\documentclass{aims}
\usepackage{amsmath}
  \usepackage{paralist}
\usepackage{graphicx}  %\usepackage{epstopdf}%This is to transfer .eps figure to .pdf figure; please compile your paper using PDFLeTex or PDFTeXify.
 \usepackage[colorlinks=true]{hyperref}
\hypersetup{urlcolor=blue, citecolor=red}
\def\currentvolume{X}
 \def\currentissue{X}
  \def\currentyear{200X}
   \def\currentmonth{XX}
    \def\ppages{X--XX}
     \def\DOI{10.3934/xx.xx.xx.xx}

\theoremstyle{definition}

\title[Spatial patterns from quasi-cycle coupling] %Use the shortened version of the full title
      {Rapidly forming, slowly evolving,  spatial patterns from quasi-cycle Mexican Hat coupling}

\author[P.E. Greenwood]{}
\author[L.M. Ward]{}

 \email{pgreenw@math.ubc.ca}
 \email{lward@psych.ubc.ca}
 
\thanks{LMW is supported by NSERC grant A9958}

\thanks{$^*$ Corresponding author. Email: lward@psych.ubc.ca,  Tel.: +1 604 822 6309, Fax: +1 604 822 6923.}

\begin{document}
\maketitle

% Enter the first author's name and address:
\centerline{\scshape Priscilla E. Greenwood}
\medskip
{\footnotesize
% please put the address of the first author
 \centerline{Department of Mathematics}
   \centerline{University of British Columbia}
   \centerline{Vancouver, BC, Canada}
} % Do not forget to end the {\footnotesize by the sign }

\medskip

\centerline{\scshape Lawrence M. Ward$^*$}
\medskip
{\footnotesize
 % please put the address of the second  and third author
 \centerline{Department of Psychology and Brain Research Centre}
   \centerline{2136 West Mall, University of British Columbia}
   \centerline{Vancouver, BC, V6T 1Z4 Canada}
}

\bigskip

% The name of the associate editor will be entered by an editorial staff
% "Communicated by the associate editor name" is not needed for special issue.
 \centerline{(Communicated by the associate editor name)}

%The abstract of your paper
\begin{abstract}
A lattice-indexed family of stochastic processes has quasi-cycle oscillations if its otherwise-damped oscillations are sustained  by noise. Such a family performs the reaction part of a discrete stochastic reaction-diffusion system when we insert a local Mexican Hat-type, difference of Gaussians, coupling on a one-dimensional and on a two-dimensional lattice. Quasi-cycles are a proposed mechanism for the production of neural oscillations, and Mexican Hat coupling is ubiquitous in the brain. Thus this combination might provide insight into the function of neural oscillations in the brain. Importantly, we study this system only in the transient case, on time intervals before saturation occurs. In one dimension, for weak coupling, we find that the phases of the coupled quasi-cycles synchronize (establish a relatively constant relationship, or phase lock) rapidly at coupling strengths lower than those required to produce spatial patterns of their amplitudes. In two dimensions the amplitude patterns form more quickly, but there remain parameter regimes in which phase synchronization patterns form without being accompanied by clear amplitude patterns. At higher coupling strengths we find patterns both of phase synchronization and of amplitude (resembling Turing patterns) corresponding to the patterns of phase synchronization. Specific properties of these patterns are controlled by the parameters of the reaction and of the Mexican Hat coupling. 
\end{abstract}
%\maketitle

%The title of your section 1
\section{Introduction}
The celebrated book of Y. Kuramoto~\cite{Kura84} begins with a description of a reaction-diffusion system ``...obtained by adding some diffusion terms to a set of (first order) ordinary differential equations." He notes that ``...the propagation of the action potential in nerves and nerve-like tissues is known to obey this type of equation." Continuing, he states that the important feature of reaction-diffusion fields, not shared by fluid dynamical systems, is that the total system can be viewed as an assembly of a large number of local systems that are all `diffusion coupled' to each other. He assumed that if one of these local systems were isolated, it would display a persistent limit cycle. ``Thus the total system may be imagined as forming a diffusion-coupled field of similar limit cycle systems." (\cite{Kura84}, page 1). A primary result was that coupling of limit cycle phases over the entire field produces synchronization of those phases over the entire field, if coupling is sufficiently strong. We extended this result to global coupling of quasi-cycles in \cite{GMWKura}. Others have extended this analysis to a wide range of different oscillating systems, e.g., \cite{Strogatz00}, including neural systems, e.g., \cite{Breakspear10, Deco16}. 
 
Here, as in \cite{GMWKura}, we address coupling of quasi-cycle systems (systems in which otherwise damped oscillations are sustained by noise) instead of limit cycle systems. But in the present case we implement a \emph{local} coupling rather than the global coupling exemplified by the Kuramoto model. In our local coupling, the systems are positioned in space so that near-by systems excite each other whereas systems farther away inhibit each other, a so-called Mexican Hat coupling. This type of local coupling has been studied for deterministic Kuramoto phase oscillators in several papers, which we will discuss in relation to our results in the Discussion section \cite{Chandra18, Heitman15, Park17, Uezu12, Uezu13}. Here we ask: can a spatial pattern of \emph{stochastic} phase synchronization result from such a local coupling? And, given this pattern in the phases, do the amplitudes of the coupled quasi-cycles exhibit a corresponding spatial pattern? We find that the answers to both of these questions is `yes' for the model we study, although there are nuances. In particular, for weak coupling the phase synchronized spatial pattern develops rapidly in the absence of a corresponding spatial amplitude pattern, whereas for stronger coupling both phase and corresponding amplitude patterns emerge.

Why would one choose to study locally-coupled quasi-cycle systems? First, there is reason to believe that quasi-cycle systems may generate brain oscillations \cite{Bressloff10, GMW14, WG16}. Second, brain oscillations are deeply related to information transmission and other brain processes \cite{Deco16}. Third, functional coupling (synchronization) of oscillations is believed to be one mechanism by which information is efficiently transmitted between brain areas \cite{Deco16}. Fourth, such couplings have been hypothesized to be Mexican-Hat-like in numerous studies at many levels of the brain, e.g., \cite{Bekesy67, Marr80, Hama04}. 

Oscillatory activity in the brain relevant to a given input likely lasts only a few hundred ms at most before changing in response to a new or changed input; the brain's oscillatory states are transient, e.g., \cite{Burns11}. Because patterns for a specific input are transient we need to understand the dynamics of the system only during bounded, in fact rather short, time intervals. We omit the usual nonlinear gain term and adjust parameters so the process stays within a bounded region of phase space during the time interval of interest. 

We shall call our model a \emph{stochastic reaction-coupling} system, to recognize that it is a local coupling of the reaction components of a stochastic system. The resulting spatial waves interact with reaction-plus-noise-generated temporal waves to form evolving spatial patterns of temporal phase ordering and closely corresponding spatial patterns of quasi-cycle amplitudes. This, we believe, is the first study of joint phase and amplitude behavior associated with a stochastic reaction-coupling system. 

In certain parameter regions reaction-diffusion equations generate Turing patterns. It is known from power spectral density computations \cite{Butler11, Butler2012, McKane14} that stochastic reaction-diffusions can generate quasi-patterns in space-time. Motivated by the existence of such psd examples, in \cite{BGW19} we explored how certain sample path properties of `stochastic neural fields,' with only a simple damping reaction term, depend on coupling strength, Mexican Hat parameters, and noise smoothing. 

Here we extend the study to evolving random fields where reaction terms produce quasi-cycles that are then coupled. An essential difference from several of our references is that we couple, not deterministic cycles, but quasi-cycles, damped oscillations sustained by noise.
 
In order to study spatial patterns of quasi-cycle phase synchronization and corresponding spatial patterns of amplitude, we compute stochastic coupling equations for the temporal phase and amplitude processes, corresponding to stochastic reaction-coupling processes expressed in rectangular coordinates, by a nontrivial application of It\^{o}'s Lemma. Simulation of the phase and amplitude evolving random fields reveals previously unseen `sample path' properties. Spatial patterns (orderings) appear rapidly among phases of the temporal oscillations, even for weak local couplings and in the absence of amplitude patterns. When coupling is strong enough, corresponding spatial patterns appear also in the amplitudes of the temporal oscillations. 

In Section \ref{app2D} we present our basic model and use It\^{o}'s Lemma to derive the stochastic differential equations for phase and amplitude components of the solution. In Section \ref{results} we describe the results of simulations in one and two spatial dimensions, and in Section \ref{discuss} we discuss these results and our model in the context of other models that involve Mexican Hat or Laplacian coupling and neural field equations.

\section{The stochastic reaction-coupling system}\label{app2D}
\subsection{Quasi-cycles}
A homogeneous stochastic reaction system that produces quasi-cycles can be written as a collection of identical stochastic diffusion processes
\begin{equation}\label{genquasi}
d\mathbb{X}_j(t)=f(\mathbb{X}_j(t))dt+g(\mathbb{X}_j(t))d\mathbb{W}_j,
\end{equation}
where $\mathbb{X}_j(t)$ has values in $\mathbb{R}^2$, $\mathbb{X}_j (t)=\big(\begin{smallmatrix} x_{1j}(t)\\x_{2j}(t)\end{smallmatrix}\big)$, and the processes $\mathbb{W}_j$ are independent $\mathbb{R}^2$ Brownian motions. We think of $j$ as indexing points in a spatial lattice in $\mathbb{R }^1$ or $\mathbb{R}^2$. We could have begun with a non-linear system such as the predator-prey example in \cite{Bax11} or a simple (SIR) epidemic model, or an excitatory-inhibitory neuron population model as will appear in Section \ref{react}, and linearized about a fixed point to obtain \eqref{genquasi}. If the deterministic system, $d\mathbb{X}_j (t)=f(\mathbb{X}_j (t))dt$, has a stable fixed point and the matrix $-\mathbb{A}_0$ obtained by linearizing around the fixed point has complex eigenvalues $-\lambda\pm i\omega$ with $0<\lambda$, the system damps to the fixed point at rate $\lambda$. If $g\ne0$ the noise in system \eqref{genquasi} causes stochastic oscillations at a distribution of frequencies, centered around $\omega$, to be maintained. These stochastic oscillations are called `quasi-cycles.' We obtain our space-time model by centering and linearizing $f$ at the fixed point, evaluating $g$ at the fixed point, to obtain $\mathbb{E}_0$, and then coupling the quasi-cycles.

\subsection{The model}\label{basic}
For each $j$ we have a linear stochastic process 
\begin{equation}\label{Kang3}
d_t\mathbb{V}_j(t)=-\mathbb{A}_0\mathbb{V}_j(t) dt + \mathbb{M} \mathbb{V}_j (t) dt+\mathbb{E}_0 d\mathbb{W}_j(t),
\end{equation}
with values in $\mathbb{R}^2$, $\mathbb{V}_j (t)=\big(\begin{smallmatrix} v_{1j}(t)\\v_{2j}(t)\end{smallmatrix}\big)$.  $\mathbb{M}$ is the coupling operator defined on a family $\xi _j (t)$ of functions of $t$ by
\begin{align}\label{operator}
\mathbb{M}\xi_{jl} (t)=\sum_l cm(j-l)\xi_j (t),
\end{align}
$c$ represents strength of coupling, and $m(j)$ is a discretization of $m(x)$, a smooth (spherically) symmetric, bounded function with support on a bounded interval, such as the Mexican Hat function \eqref{diffG}. $\mathbb{M}$ represents a local spatial operator, here the difference-of-Gaussians (Mexican Hat) operator or its discrete approximation. In Kuramoto's field of coupled limit cycle phases, and in many other applications of that model, the operator $\mathbb{M}$ is, instead, the Laplacian, or the discretized Laplacian.

The noise, denoted $d\mathbb{W}_j (t)$ is standard temporal Gaussian noise with independent components and is independent for each $j$. With the coefficient $\mathbb{E}_0$ the noise term has temporal covariance matrix $\mathbb{B}_0=\mathbb{E}_0 \mathbb{E}_0^\top$. Space is wrapped to avoid boundary conditions. It is also interesting to consider spatially smoothed noise, as in \cite{BGW19}, but we do not do this here.

Although the separate systems in \eqref{Kang3} without the middle coupling term and without noise would damp to a fixed point, for $c$ above a critical value, when the systems in \eqref{Kang3} are coupled by $\mathbb{M}$ the resulting system is unstable. In several models involving similar equations, a nonlinear "squashing" functional, often the logistic, operates on the coupling term to keep the entire system bounded. Instead, we adjusted our parameter values, particularly those of the Mexican Hat operator, to keep the system \eqref{Kang3} in the linear region, and stochastically bounded, on bounded time intervals.

\subsection{Reaction term of coupled quasi-cycle model}\label{react}
For the reaction term in \eqref{Kang3} we have in mind a family of models often considered in mathematical neuroscience where populations of excitatory (E) and inhibitory (I) neurons interact according to a scheme that is an example of our basic model \eqref{Kang3}. Suppose we have a family of $N$ excitatory-inhibitory subpopulation models indexed by $j = 1, 2, ...N$, as in \cite{GMWKura,Kang10}. For each $j$ the model \eqref{Kang3} without coupling will be
\begin{align}\label{Kang1}
\tau_E dV_{E}(t)&=(-V_{E}(t)+S_{EE}V_{E}(t)-S_{EI}V_{I}(t)) dt + \sigma_{E} dW_{E}(t)\notag\\
\tau_I dV_{I}(t)&=(-V_{I}(t)-S_{II}V_{I}(t)+S_{IE}V_{E}(t)) dt + \sigma_{I} dW_{I}(t).
\end{align}
In (\ref{Kang1}) $W_E,W_I$ are independent, standard Brownian motions. $S_{EE}, S_{II}, S_{IE}, S_{EI} \ge 0$ are constants representing the efficacies of excitatory or inhibitory synaptic connections to post-synaptic neurons within each separate population, as indicated by the notation, with $S_{IE}$ representing input to inhibitory from excitatory neurons. These parameters, along with the time constants, $\tau_E,\tau_I$, determine the oscillatory behaviour of the system and in particular its dominant frequency of oscillation. The amplitudes of the Brownian motions, $\sigma_E,\sigma_I$, determine the amplitudes of the oscillations that are sustained when they are non-zero. When \eqref{Kang1} is expressed in the notation of \eqref{Kang3} we have $-\mathbb{A}_0=\bigg(\begin{smallmatrix} (1-S_{EE})/\tau_E& S_{EI}/\tau_E\\-S_{IE}/\tau_I& (1+S_{II})/\tau_I\end{smallmatrix}\bigg)$. The dominant frequency of oscillation, $\omega$, arises from the complex eigenvalues, $-\lambda\pm i\omega$, of $\mathbb{A}_0$ (see \cite{GMW14} for an extended discussion of this model). For simulation we chose a parameter set (see Table 1 in Section \ref{1D1}) where the oscillation is narrow-band and thus has a distinct phase even though it arises from a stochastic process. 

An essential point is that without the noise, i.e., with $\sigma_E=\sigma_I=0$, the temporal oscillations damp to zero at rate $\lambda$. With small noise the oscillations are sustained and are called quasi-cycles.  

\subsection{Neural motivation}\label{neuralfield}
\begin{figure}[!ht]
\begin{center}
\includegraphics[width=5in]{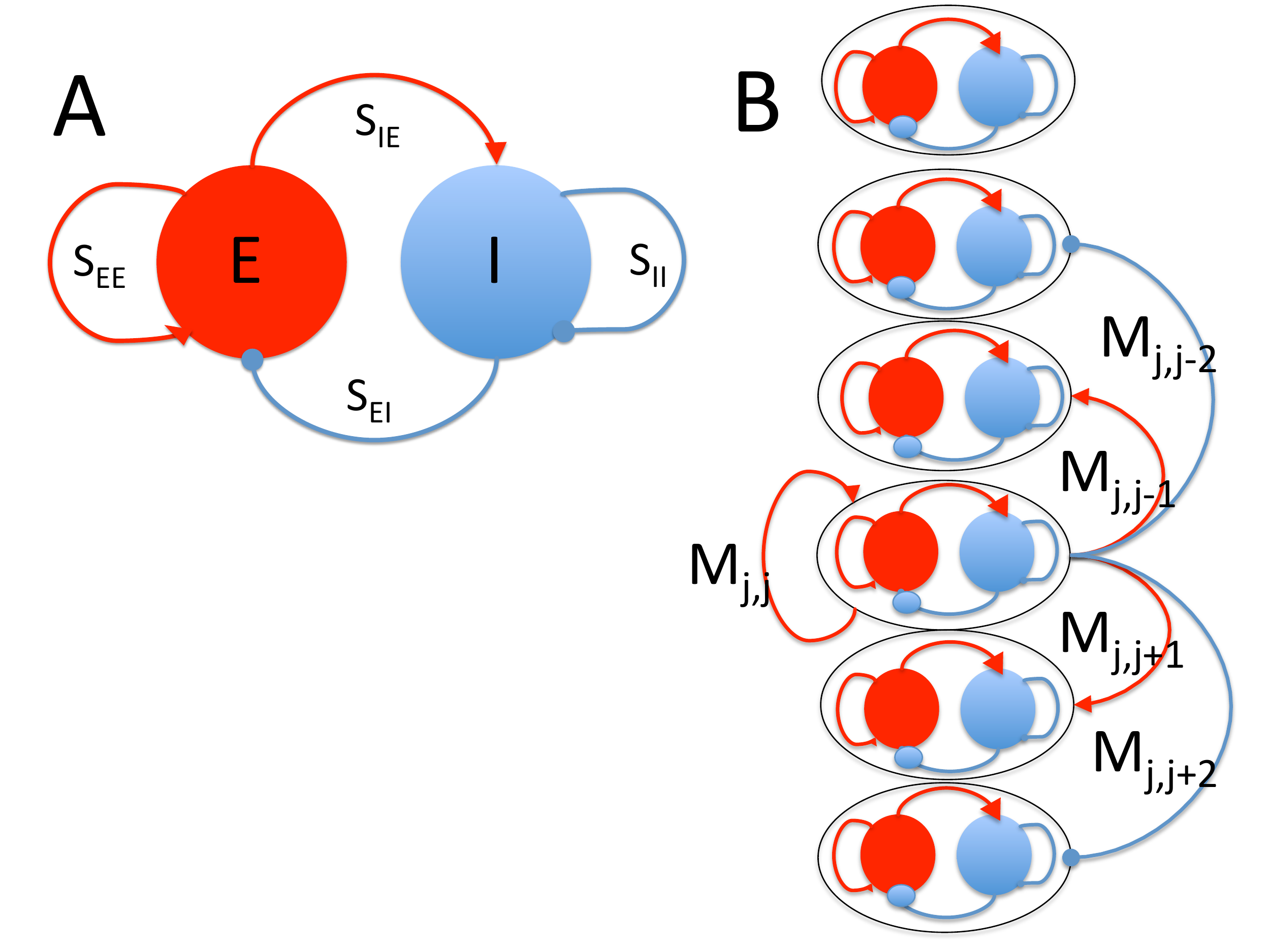} 
\end{center}
\caption{A. Diagram of excitatory (red lines) and inhibitory (blue lines) connections between excitatory (E) and inhibitory (I) collections of neurons that, with appropriate parameter values in a specific form of \eqref{Kang3}, generate quasi-cycle oscillations.  B. Microcircuits as in A connected by a minimal Mexican Hat coupling, as described in Appendix B. Again, red lines represent excitatory connections and blue lines represent inhibitory connections. In the simulations to be described later, there are 128 copies of the microcircuit depicted in A, arranged in a ring, and the Mexican Hat coupling extended over 31 microcircuits, 15 to each side of \emph{each} microcircuit, rather than over only five as depicted in the figure.} 
\label{FigEI}
\end{figure}

As a motivating example let us interpret \eqref{Kang1} as comprising an oscillatory system made up of a population of excitatory and inhibitory neurons, and characterized by a particular dominant frequency. Figure \ref{FigEI}A displays a schematic of this model. When a system of such `microcircuits' is functionally coupled by an operator such as the Mexican Hat operator, $\mathbb{M}$ in \eqref{Kang3}, the extended model schematically depicted in Figure \ref{FigEI}B results. Here each microcircuit is functionally coupled to its nearby neighbors by excitatory connections, and to some of its more distant neighbors by inhibitory connections. Although not meant in this paper to represent actual neural circuitry in any particular brain area, this scheme is similar to those proposed for, e.g., feature detectors in later visual areas \cite{Brincat2006}, memory representations \cite{Salinas96}, or gnostic units in association cortex \cite{Walley1973}, among others. It should be noted that several different types of local neural connectivity could result in a functional scheme like the one assumed here. For example, for a given distribution of excitatory connections, the spatial distribution of inhibitory connections could be narrower if the inhibition is faster than the excitation, but broader if there is a significant population of slower excitatory synapses (e.g., NMDA-type) \cite{Kang03}. A broader distribution of inhibition could be mediated by basket-type GABAergic neurons \cite{Buzas01}. 

System \eqref{Kang1} is an example of the local linear micro-structure without the coupling term containing $\mathbb{M}$. Inserting $\mathbb{M}$, as in \eqref{Kang3}, results in two levels of E-I type interactions. We wish to emphasize that the Mexican Hat coupling we introduce in \eqref{Kang3} represents functional connections, not specific neural implementation of those functional connections. We do not specify exactly how the excitatory and inhibitory elements of the microcircuits participate, if at all, in the excitatory and inhibitory connections at the network level.

\subsection{Stochastic phase and amplitude equations}
In order to produce stochastic paths corresponding to the model \eqref{Kang3}, specified, for example, by the reaction term(s) \eqref{Kang1}, we change variables and compute corresponding phase and amplitude equations, simplified by beginning with the matrix $\mathbb{A}_0$ changed to normal form. Let $\mathbb{Q}$ be a 2x2 matrix such that
\begin{equation}\label{A}
\mathbb{Q}^{-1}(-\mathbb{A}_0)\mathbb{Q} = \begin{pmatrix}
									-\lambda&  \omega\\
									-\omega&  -\lambda
									\end{pmatrix}
									:= \mathbb{A}.
\end{equation}
Such a matrix is
\begin{equation}
\mathbb{Q}=\begin{pmatrix}
				-\omega& \lambda-\mathbb{A}_{011}\\
				0&  -\mathbb{A}_{021}
				\end{pmatrix}.
\end{equation}
We change variables in \eqref{Kang3}, putting 
\begin{align}\label{Y}
\mathbb{Y}_j (t)=\mathbb{Q}^{-1}\mathbb{V}_j (t),
\end{align}
to obtain, because $\mathbb{Q}$ commutes with the operator $\mathbb{M}$,
\begin{align}\label{dY}
d_t \mathbb{Y}_j (t)=\mathbb{A}\mathbb{Y}_j (t)dt+\mathbb{M}\mathbb{Y}_j (t)dt+\mathbb{E}d\mathbb{W}_j {(t)},
\end{align}
where $\mathbb{E}:=\mathbb{Q}^{-1}\mathbb{E}_0$. For simplicity in our computations we take the covariance matrix $\mathbb{E}=\mathbb{I}$. We will regard \eqref{dY} as our model, with $Y_j (t)=\big(\begin{smallmatrix} y_{1j}(t)\\y_{2j}(t)\end{smallmatrix}\big)$. Using It\^o's Lemma we obtain the following stochastic equations for the space-time processes $\theta_j=\arctan(y_{2j}/y_{1j})$ and $Z_j=(y_{1j}^2+y_{2j}^2)^{1/2}$ (see Appendices A and B):
\begin{align}\label{Mexphase}
d\theta_{j}= \omega dt
+\Bigg[\sum_{l=1}^{N} \frac{Z_l (t)}{Z_j (t)} \mathbb{M}_{jl} \sin (\theta_j (t)-\theta_l (t))\Bigg] dt
+\frac{db_j(t)}{Z_j(t)},
\end{align}
where $\omega$ is the frequency in \eqref{A}, and
\begin{align}\label{MexZ}
dZ_{j}=
\left(\frac{1}{2Z_{j}(t)}-\lambda Z_{j}(t)\right) dt
+\bigg[\sum_{l=1}^{N} \mathbb{M}_{jl} Z_{l}cos(\theta_j (t)-\theta_l (t))\bigg] dt
+dW_{j}(t),
\end{align}
where $b_j (t)$ is Brownian motion on the unit circle, and $\mathbb{M}_{jl}$ represents the Mexican Hat coupling, $\mathbb{M}$, acting over a specific range of the spatial lattice, and 0 outside that range. 

Notice that whenever we have \eqref{dY} the stochastic change of variables will result in \eqref{Mexphase} and \eqref{MexZ}. As \eqref{dY} will result from normalization of a wide range of coupled reaction systems, including many arising in population dynamics, epidemiology, or a system like that of \cite{Butler2012}, this approach, writing \eqref{dY} as an evolving random field in polar coordinates, has wide generality. Appendix A suggests that if $\mathbb{M}$ is a discrete Laplacian, instead of a Mexican Hat, we will see similar results in simulations of sample paths.

Let us pause to consider the dynamics expressed by \eqref{Mexphase} and \eqref{MexZ}. If $Z_j (t)$ were constant in $j$ and $t$, \eqref{MexZ} would look much like Kuramoto coupling \cite{Kura84}. The difference is that in Kuramoto's case $\mathbb{M}_{jl}$ is a constant $c$ for all $j,l$, expressing all-to-all coupling. It will turn out that $\theta_j (t)$ approaches a bilinear function. The effect of the local coupling does not stay local in space, but spreads.

In \eqref{MexZ} the Mexican Hat coupling of each pair of amplitudes, $Z_j, Z_l$, is through the cosine of the difference between their corresponding phases, $\theta_j, \theta_l$: $\mathbb{M}_{jl}Z_j cos(\theta_j (t)-\theta_l (t))$. Where the phases are similar, i.e., phase differences near zero, the coupling has the largest effect on the amplitude, because there the cosine is near 1 or -1. It will turn out that the coupling, when sufficiently strong, produces a pattern in the amplitude, $Z_j$, that reflects the spatial ordering in the phase processes.

We wish to emphasize here that we will refer to a situation in which phases $\theta_j, \theta_l$ maintain a relatively consistent ordering as they progress over the range $-\pi$ to $\pi$ to $-\pi$ and so forth, no matter what that difference is, as `ordering' of phase, similar to the use of the words `synchronization' or `phase locking' in other contexts \cite{Lachaux99}. We note that in theoretical neuroscience specific phase relationships between oscillating neural systems have been proposed to facilitate information transmission between them \cite{Fries05, Sejnow10}. Our usage of the experession `phase ordering' is meant to be consistent with this usage. 

In spite of the fact that the uncoupled individual $\mathbb{R}^2-$valued processes, i.e. $\mathbb{M}_{jl}=0$ in \eqref{Mexphase} \eqref{MexZ}, produce quasi-cycles, after coupling by a Mexican Hat operator sufficient to produce unstable states the marginal processes do not do so. To be explicit, the system \eqref{Mexphase} \eqref{MexZ} is unstable for the Mexican Hat parameters we study: $Z_j$ generally increases exponentially for all $j$ for long time intervals whenever $c\ge 5$. This in turn quenches the phase noise because of the final term in \eqref{Mexphase}, resulting in a deterministic rotation of ever-increasing amplitude. Before this would occur in a neural system the firing rate would saturate at its maximum, limited by the duration of a spike and the refractory period to about 200-500 spikes/sec. Here we study the transient response (with one exception) only where the $Z_j$ remain relatively small and bounded and the neural response would be in a functional range below saturation. This is the case most likely relevant to neural systems. We expand on this point in the Discussion.

\section{Numerical Results}\label{results}
There follows our numerical study of the properties of the discrete Mexican-Hat-coupled system of quasi-cycle oscillators. Copies of processes \eqref{Kang1}, indexed by $j$ denoting location on a discrete lattice in $\mathbb{R}^1$ or $\mathbb{R}^2$, coupled by the Mexican Hat operator as in \eqref{dY} and expressed in polar coordinates in \eqref{Mexphase}, \eqref{MexZ}, comprise our reaction-coupling dynamics. In \eqref{Mexphase}, \eqref{MexZ}, the reaction and noise terms produce quasi-cycles whose phases and amplitudes are then coupled by the Mexican Hat operator. We varied the Mexican Hat operator in both one and two spatial dimensions. We are interested in the question: how do spatial waves produced by local coupling combine with point-based temporal quasi-cycles to form an evolving random field of quasi-patterns? We display plots of representative sample paths of the systems $\mathbb{V}(t,x)$ defined by \eqref{Kang3} via the discretized polar systems \eqref{Mexphase}, \eqref{MexZ}, with specific parameters, in one spatial dimension, and fixed-time plots in two spatial dimensions for \eqref{Mexphase} and \eqref{MexZ}. 

We solved numerically the relevant SDEs, with parameters given in Table \ref{Table1}, using the Euler-Maruyama iterative method \cite{Kloeden}. We varied the coupling strength between the systems to generate the spatial patterns displayed. In one spatial dimension we simulated the local coupling of 128 quasi-cycle processes indexed by $j$ with periodic boundary. So the spatial variable, $j=1,....,128,$ can be considered to form a loop or ring. In two dimensions we simulated a 100x100 (=10,000 processes) lattice with a neutral (no coupling) boundary. The basic procedure was the same for all computations. A discretized difference-of-Gaussians operator was our Mexican Hat operator (see \cite{BGW19}). The operator comprised a 31 point vector, indexed by $j$, in the 1-D simulations, and comprised a 21x21 point matrix, indexed by $j,l$ in the 2-D simulations. The time step was small, 0.00005 seconds, in order to avoid problems with stiff solutions. Noise increments were always i.i.d. standard Gaussian multiplied by the square root of the time step.

\begin{table}[h!]
\caption{Parameters used in simulations and for figures.}\label{Table1}
\begin{center}
\begin{tabular}{|c|c|c|}
\hline
Variable & Value & Units\\
\hline
\hline
$S_{II}$ & 0.1 & dimensionless\\
\hline
$S_{EE}$ & 1.5 & dimensionless\\
\hline
$S_{EI}$ & 1.0 & dimensionless\\
\hline
$S_{IE}$ & 4.0 & dimensionless\\
\hline
$\tau_E$ & 0.003 & seconds\\
\hline
$\tau_I$ & 0.006 & seconds\\
\hline
$\lambda$ & 8.333 & 1/seconds\\
\hline
$\omega$ & 437.72  or 69.66 & radians per second or Hz\\
\hline
$\lambda/\omega$ &$0.019$ & dimensionless\\
\hline
$\Delta t$ & 0.00005 & seconds\\
\hline
\end{tabular}
\end{center}
\end{table}

\subsection{Stochastic reaction-coupling field in one spatial dimension}\label{1D1}
We considered 128 quasi-cycle-generating processes, with noise, before coupling, as described in Section \ref{app2D}, arranged in a ring (periodic boundary condition), all oscillating at the resonant frequency $\omega=437.72$ rad/s, similar to the system we studied for the Kuramoto model \cite{GMWKura}. For convenience, we began each realization with the phases of the 128 systems distributed randomly between $-\pi$ and $\pi$, and the amplitudes distributed as 0.5 plus 0.1 times a sample from the uniform distribution on [0,1]. This ensured that any synchronization of phases, or spatial patterns of amplitudes, would be produced by the Mexican Hat operator and not because the processes were started in a synchronized or patterned state.

We employed as the coupling operator \cite{Murray89} a (truncated), discretized, difference of Gaussian functions (Mexican Hat), written in continuous space variable, $x$, as
\begin{align}\label{diffG}
m(x)=b_1 \exp \bigg[-\bigg(\frac{x}{d_1}\bigg)^2\bigg]-b_2 \exp \bigg[-\bigg(\frac{x}{d_2}\bigg)^2\bigg],  b_1>b_2, d_2>d_1. 
\end{align}
In \eqref{diffG} $b_1 ,b_2$ are the heights of the Gaussian functions at $x=0$, and $d_1,d_2$ are their dispersions. We used parameters $b_2= d_1=1.0$ and $b_1, d_2$ with various values as indicated in our figures. The values of the latter two parameters determine the dominant wave number of the spatial pattern produced in our system \cite{Murray89}. In the discrete version used in computation, the Mexican Hat kernel is represented by numbers $\mathbb{M}_{jl}=chm(j)$, where the constant $c\ge0$ is termed the `coupling strength,'  and $j$ varies (in steps of 1) from $x=$ -3 to 3, the effective range of our Mexican Hat, in 30 steps of $h=0.2$ in x-space, thus making the ring of length $L=nh=25.6$. Outside the region of the Mexican Hat ($j-l>15$), the $\mathbb{M}_{jl}$ of \eqref{Mexphase} and \eqref{MexZ} were set to zero. In this way a range of different strengths of coupling can be generated for a given set of parameters $b_1, d_2$.  Multiplying \eqref{diffG} by $c$ in \eqref{operator} changes the heights of the two Gaussian functions composing $m(j)$ for a given $b_1, d_2$ without changing their dispersions: $chm(j) = ch b_1 \exp(-(j/d_1)^2) - ch b_2 \exp(-(j/d_2)^2).$ Multiplying \eqref{diffG} by $c$ also multiplies its Fourier transform by $c$, which determines the effectiveness of the coupling in generating spatial patterns (cf. \cite{BGW19}). Just how the particular $\mathbb{M}$ operator affects both the speed and the type of pattern development, will be explored later in this section. 

In order to illustrate spatial patterns, for each set of parameter values we display a representative realization of the evolving random field consisting of the paths of all 128 processes, both phase and amplitude, with the ring indexed by integers. We also display, for some of those realizations, the amplitudes of the Fast Fourier Transform (FFT) components of the spatial pattern of amplitudes, $Z_j (t,x)$ as a stochastic process in $t$. For the FFT amplitudes we coarse-grained time, considering 500-iteration time blocks: 1-500, 750-1250, 1750-2250, ..., 9501-10000, and then averaged the amplitude of each process over the 500-iteration block and computed the FFT on the resulting spatial array.

\subsubsection{Simulation results in one spatial dimension}
\begin{figure}[!ht]
\begin{center}
\includegraphics[width=5in]{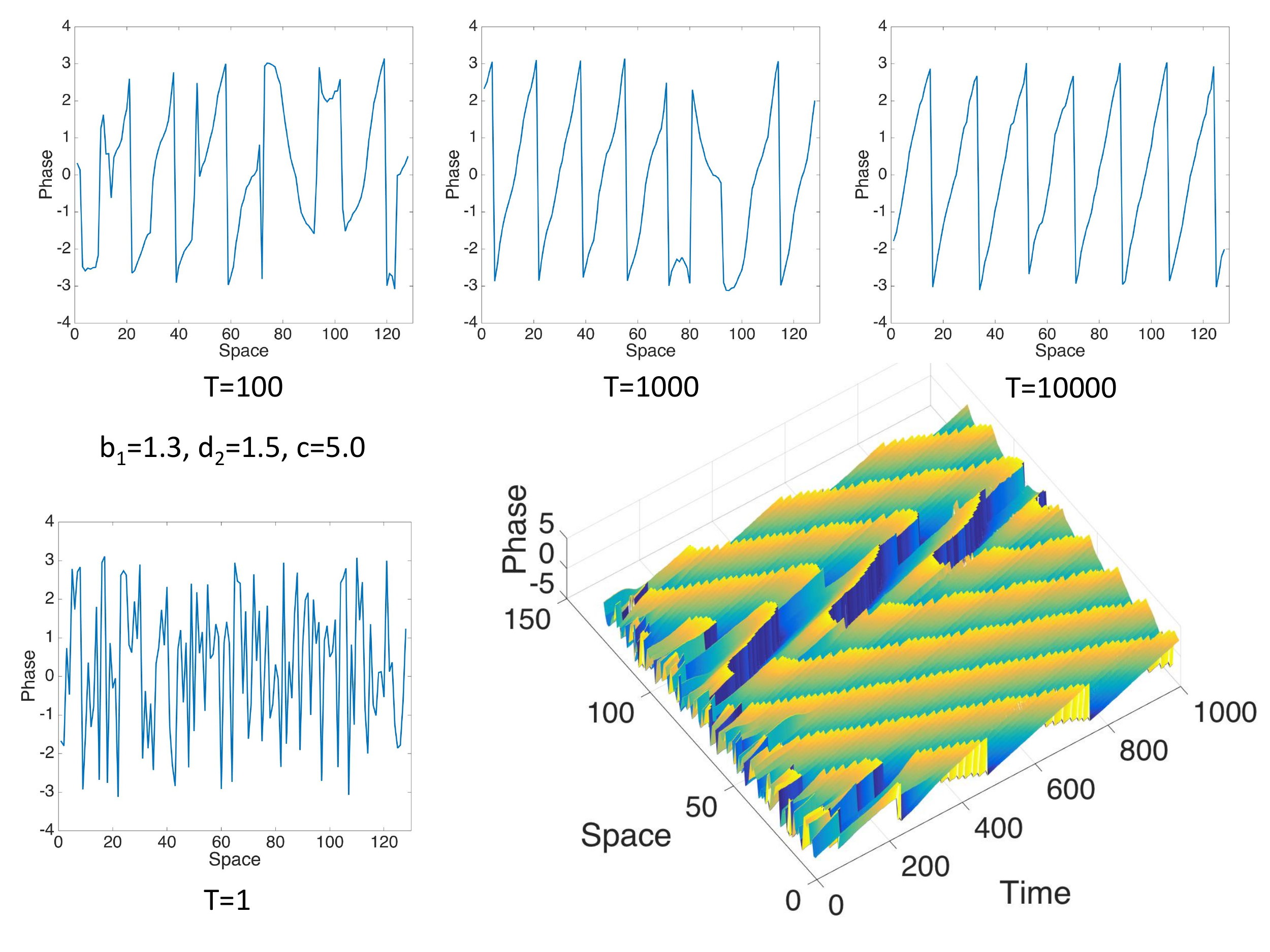} 
\end{center}
\caption{Phase dynamics for MH-coupled phase equation \eqref{Mexphase} on a ring.  Parameters were $b_1=1.3, d_2=1.5, c=5$ and those in Table 1. } 
\label{Figb1p3d1p5theta}
\end{figure}

\begin{figure}[!ht]
\begin{center}
\includegraphics[width=5in]{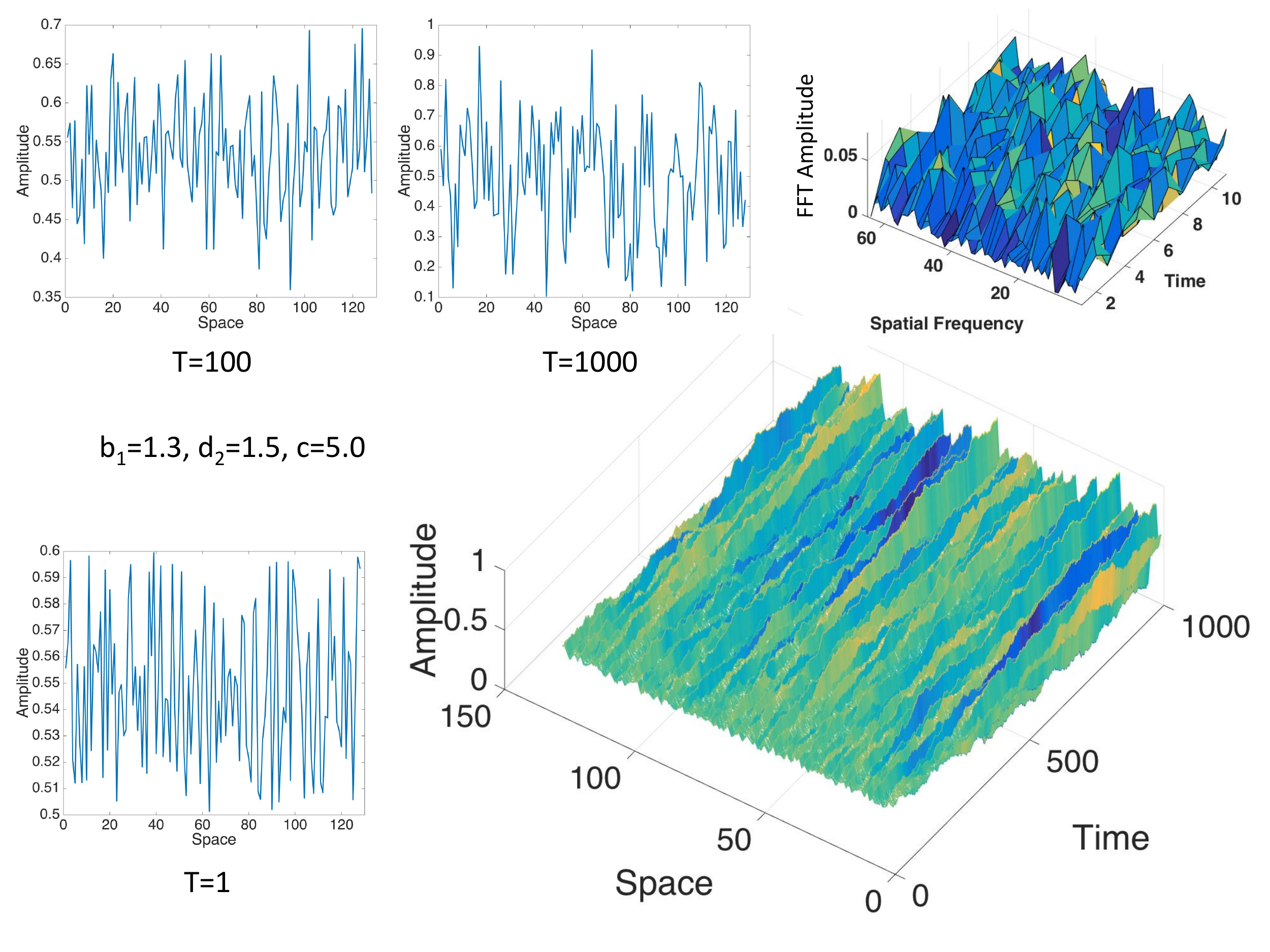} 
\end{center}
\caption{Amplitude dynamics for MH-coupled amplitude equation \eqref{MexZ} on a ring. Parameters were $b_1=1.3, d_2=1.5, c=5$ and those in Table 1, the same as in Figure \ref{Figb1p3d1p5theta}. } 
\label{Figb1p3d1p5Z}
\end{figure}

A novel result evident in Figures \ref{Figb1p3d1p5theta} and \ref{Figb1p3d1p5Z} is that ordering of phases among the component processes occurs rapidly after a random onset and at coupling strengths, $c$, of the Mexican Hat operator such that the overall amplitude is increasing, i.e., $c=5$, but still well below those required to produce spatial pattern in the amplitudes. Figures \ref{Figb1p3d1p5theta} and \ref{Figb1p3d1p5Z} display the results of an illustrative simulation of \eqref{Mexphase}, \eqref{MexZ} for $b_1=1.3, d_2=1.5, c=5$. In Figure \ref{Figb1p3d1p5theta} it can be seen that a spatial ordering of phases (or phase locking) is already well-established by iteration 100 (T=100) for this weak local coupling. That is, locations that are a particular distance apart in the ring maintain a relatively constant relationship between their phases even as the phase progresses (and is wrapped from $\pi$ to $-\pi$). Every approximately 18 locations in the ring the pattern repeats so that there are approximately 7 cycles in the spatial pattern of the phases.
 
Note that because the oscillatory phases are wrapped to the region $-\pi$ to $\pi$, discontinuities in the temporal progression of each process occur at times when the phase adjustment of $-2\pi$ occurs. The phase ordering continues to evolve throughout 10,000 iterations of the processes. Indeed, once the spatial ordering becomes stable, it will precess around the ring of processes at approximately 70 Hz, the temporal frequency of the individual processes. The limiting ordering as $t$ becomes large is described in Section \ref{discuss}.  

Figure \ref{Figb1p3d1p5Z} shows that for the same parameter values as in Figure \ref{Figb1p3d1p5theta} no spatial pattern at all develops in the amplitudes of the quasi-cycles during the same time interval, and indeed none appears over the full 10,000 iterations. The ridges in the space-time plot of the amplitudes do not indicate spatial pattern, but simple continuity, and slow growth in time, of the process amplitude, $Z_j (t,x)$ and the random initial condition at each spatial location. Thus, a higher amplitude initial value tends to be maintained in time, as does a lower amplitude. In other words, the amplitudes features remain localized. This is corroborated by the FFT amplitude plot, which shows no peaks of power at any frequency at any time interval in the iterations.  

This result, rapid ordering of phases in the absence of amplitude patterns, holds for a wide range of other combinations of values for $b_1, d_2, c$ (not shown). Typically, when $b_1, d_2$ are larger, $c$ must be smaller for a similar result to obtain. The rapid development of a spatial ordering of phases with very weak local coupling in a field of quasi-cycle oscillators is not expected from the work of Murray \cite{Murray89} or indeed from any other work with Mexican Hat operators, of which we are aware, that focuses on spatial patterns of \emph{amplitude}. Some similar effects of weak local Mexican Hat coupling of limit-cycle oscillators do occur in deterministic scenarios, however; we discuss these in the Discussion section. This result reminds us of a property of relaxation oscillators, which, when coupled, synchronize their phases rapidly without affecting each other's amplitudes \cite{vanderpol26, Somers93}. 

\begin{figure}[!ht]
\begin{center}
\includegraphics[width=5in]{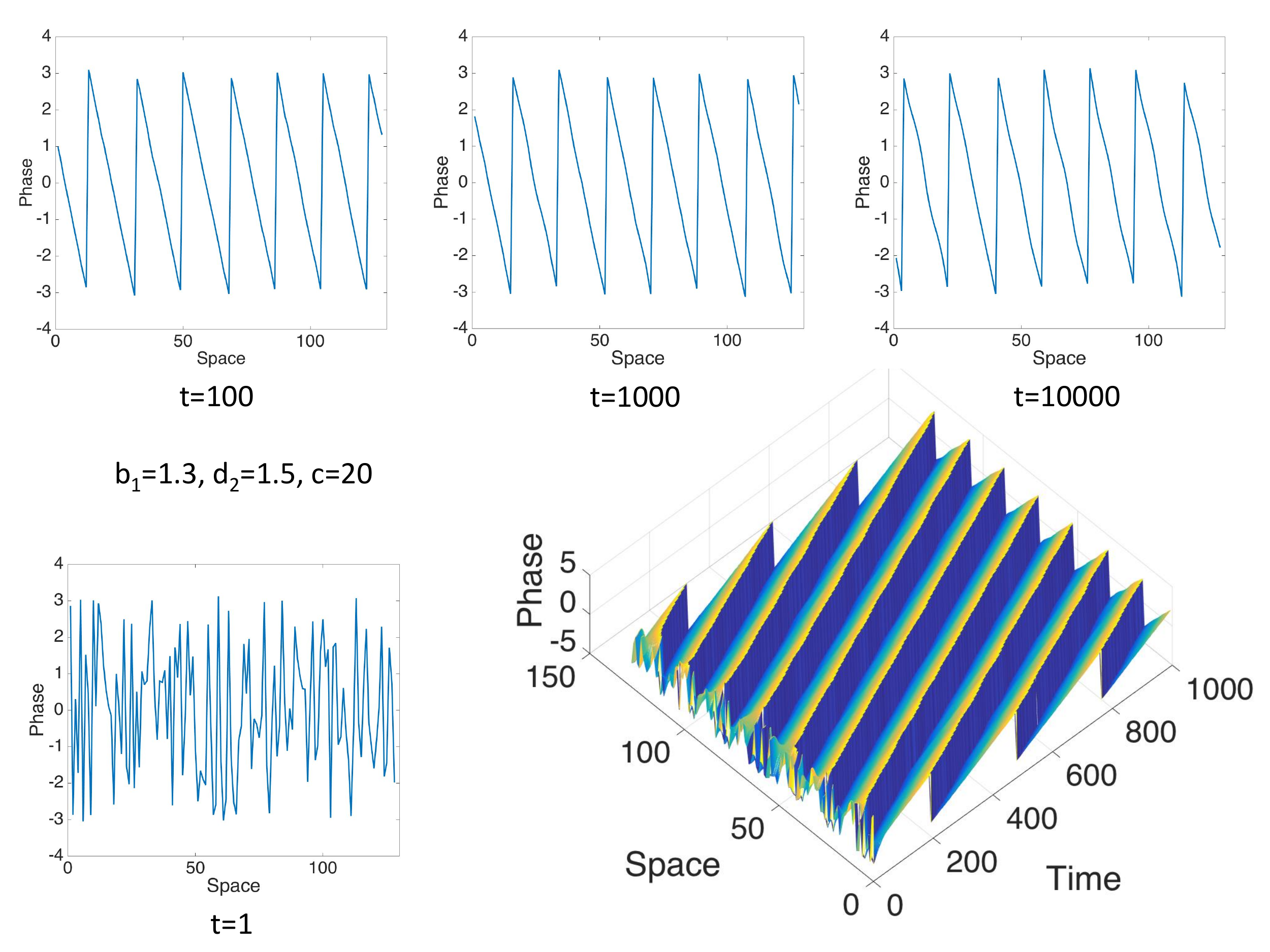} 
\end{center}
\caption{Phase dynamics for MH-coupled phase equation \eqref{Mexphase} on a ring.  Parameters were $b_1=1.3, d_2=1.5, c=20$ and those in Table 1. } 
\label{Figb1p3d1p5c4theta}
\end{figure}

\begin{figure}[!ht]
\begin{center}
\includegraphics[width=5in]{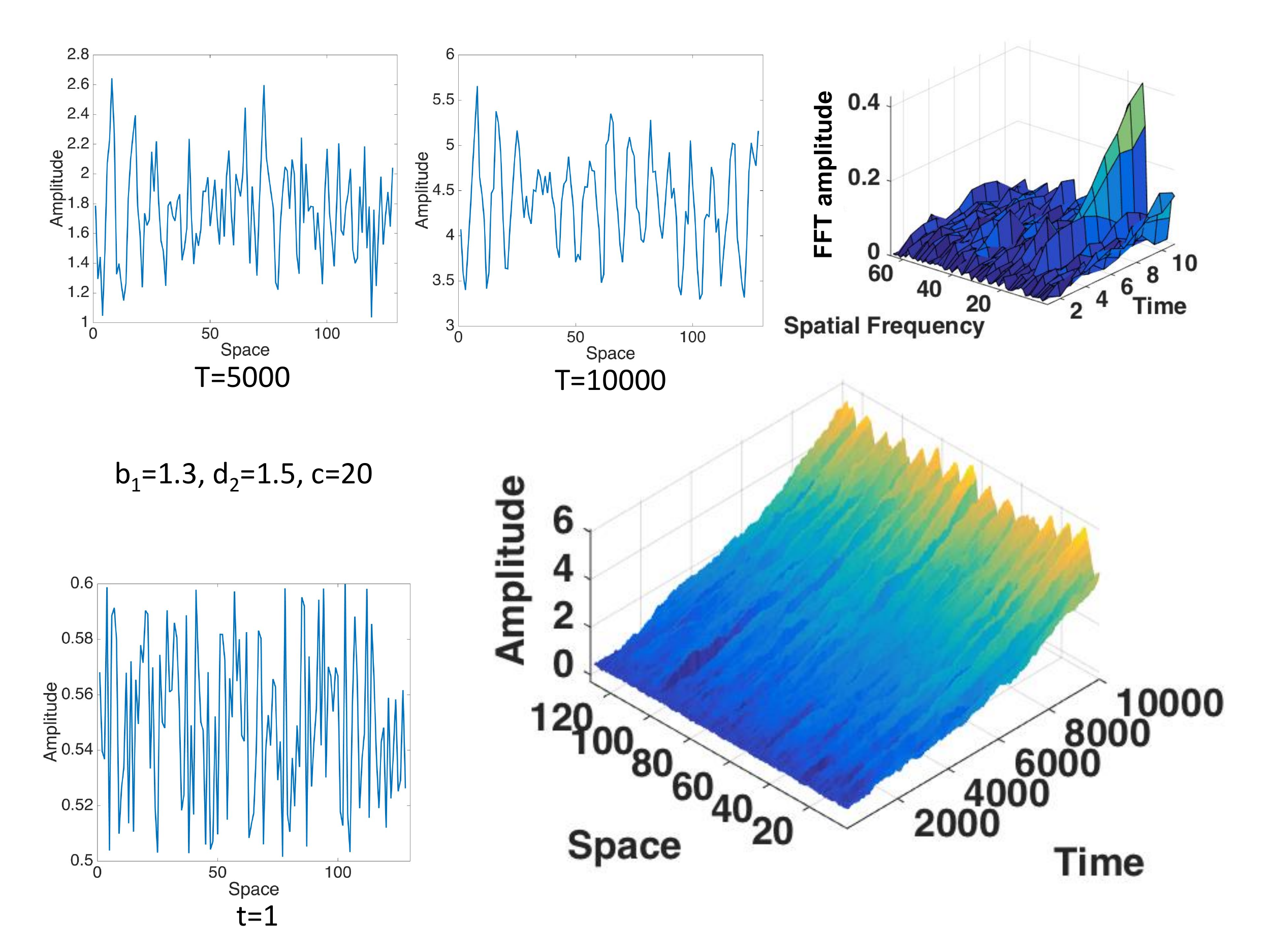} 
\end{center}
\caption{Amplitude dynamics for MH-coupled amplitude equation \eqref{MexZ} on a ring. Parameters were $b_1=1.3, d_2=1.5, c=20$ and those in Table 1, the same as in Figure \ref{Figb1p3d1p5c4theta}. } 
\label{Figb1p3d1p5c4Z}
\end{figure}

Figures \ref{Figb1p3d1p5c4theta} and \ref{Figb1p3d1p5c4Z} display the results of an illustrative simulation of the 1D model for the same $b_1=1.3, d_2=1.5$ as in Figures \ref{Figb1p3d1p5theta} and \ref{Figb1p3d1p5Z} but with stronger coupling $c=20$. These figures show the existence of spatial patterns in both phase and amplitude. The phase ordering in Figure \ref{Figb1p3d1p5c4theta} is present from an early time point and persists with little change throughout the simulation. In contrast, the amplitude pattern in Figure  \ref{Figb1p3d1p5c4Z} takes some time to develop, even with the stronger coupling. For this reason the stochastic paths of the amplitudes are shown for the entire 10,000 time steps, and the paths have been captured at somewhat different time points. The FFT amplitude plot in Figure \ref{Figb1p3d1p5c4Z} shows that the spatial amplitude pattern begins to emerge around iteration 4000, and the $t=5000$ plot shows that it is partially developed by that time. By $t=10,000$ the amplitude pattern is fully established but still somewhat noisy. Note that amplitude is  increasing for this stronger coupling somewhaqt faster than in Figure \ref{Figb1p3d1p5Z}. This amplitude increase continues indefinitely because the system is unstable, so that the system eventually leaves the region of quasi-cycles (and linearity), and deterministic rotation dominates. 

The phase and amplitude patterns Figures \ref{Figb1p3d1p5c4theta} and \ref{Figb1p3d1p5c4Z} are related. In general, amplitudes are relatively larger where the processes are approximately in ordered phase.  (Compare the phase patterns in Figure \ref{Figb1p3d1p5c4theta} with the ones in Figure \ref{Figb1p3d1p5theta}, which resulted from weaker coupling). The number of spatial cycles on the ring of amplitude processes is exactly twice the number produced in the spatial phase. This is true in general because each spatial cycle in phase results in two maxima in amplitude.

\subsection{Stochastic reaction-coupling field in two spatial dimensions}
We simulated processes \eqref{Mexphase} and \eqref{MexZ} as in Section \ref{1D1} on a 100x100 lattice (10,000 processes in all). The processes and the 2-D Mexican Hat operator all had the same parameter ranges as in the 1-D case. The outer boundary of the discrete 2D MH operator was made as circular as the discretization allowed. The two axes through the center of the operator covered 21 spatial locations. We simulated the 10,000, locally-coupled (except for a boundary band 1/2 the width of the operator around the outside of the lattice), stochastic phase and amplitude processes for 2000 iterations and examined the spatial pattern of phases and amplitudes at various points during the runs. Note again that the spatial patterns we see are those comprised of oscillations in time, represented separately by their phases and amplitudes. 

\begin{figure}[!ht]
\begin{center}
\includegraphics[width=5.0in]{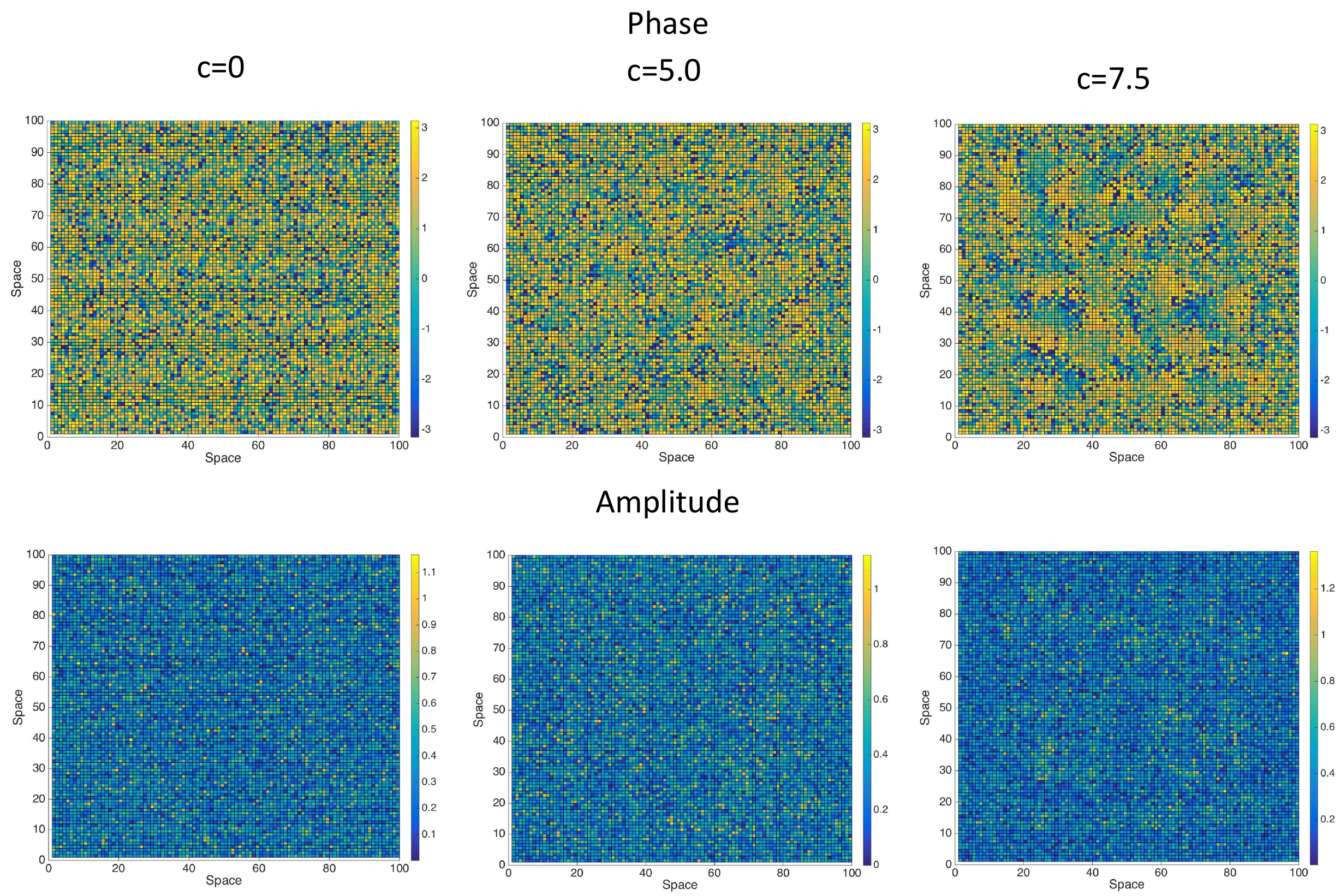} 
\end{center}
\caption{Values of phase (top row) and amplitude (bottom row) after 2000 iterations of 10,000 MH-coupled processes running on a lattice. Parameters were $b_1=1.3, d_2=1.5, c=0, 5.0, 7.5$ and those in Table 1. Neutral boundary comprised the 10 outside processes on each side, so only the 6400 processes inside this boundary were coupled via the 2-D Mexican Hat.} 
\label{Fig2Drapidsynch}
\end{figure}

\subsubsection{Simulation results in two spatial dimensions}
First we examined whether a result, similar to the 1D case, of a rapidly developing, slowly evolving spatial pattern of phase ordering, accompanied by little or no spatial pattern in the amplitudes, would result from some combination of Mexican Hat parameters and weak coupling strength in the 2-D case. A difference between our 1-D and 2-D simulations is in the boundary conditions: for the 2-D simulation the boundary with the non-coupled region was abrupt. This is actually similar to the boundaries between functional and structural areas in the brain, but introduces a boundary condition that might affect the results. We used a `neutral' boundary, in which the processes were not coupled via the Mexican Hat operator, as described earlier, for the 2-D simulations. Figure \ref{Fig2Drapidsynch} shows one such simulation at iteration 2000, with $b_1=1.3, d_2=1.5, c=0, 1.0, 1.5, noise\,SD=1.0$. Recall that for the 1-D simulations the spatial pattern of phase ordering was already well-established by iteration 1000. We expect to see a similar pattern here if the coupling is creating spatial patterns. Clearly, when coupling is absent, $c=0$, there is no spatial pattern apparent in either the phases or the amplitudes. For $c=5.0$, however, there is a spatial ordering in the phases, albeit rather weak, but no pattern in the amplitudes. When coupling is increased to $c=7.5$ the ordering in the phases is more apparent, but the pattern in the amplitudes is only weakly present. This result is similar to that found in the 1-D simulations, although there the amplitude patterns were simply absent even when strong phase ordering was evident. 

\begin{figure}[!ht]
\begin{center}
\includegraphics[width=5in]{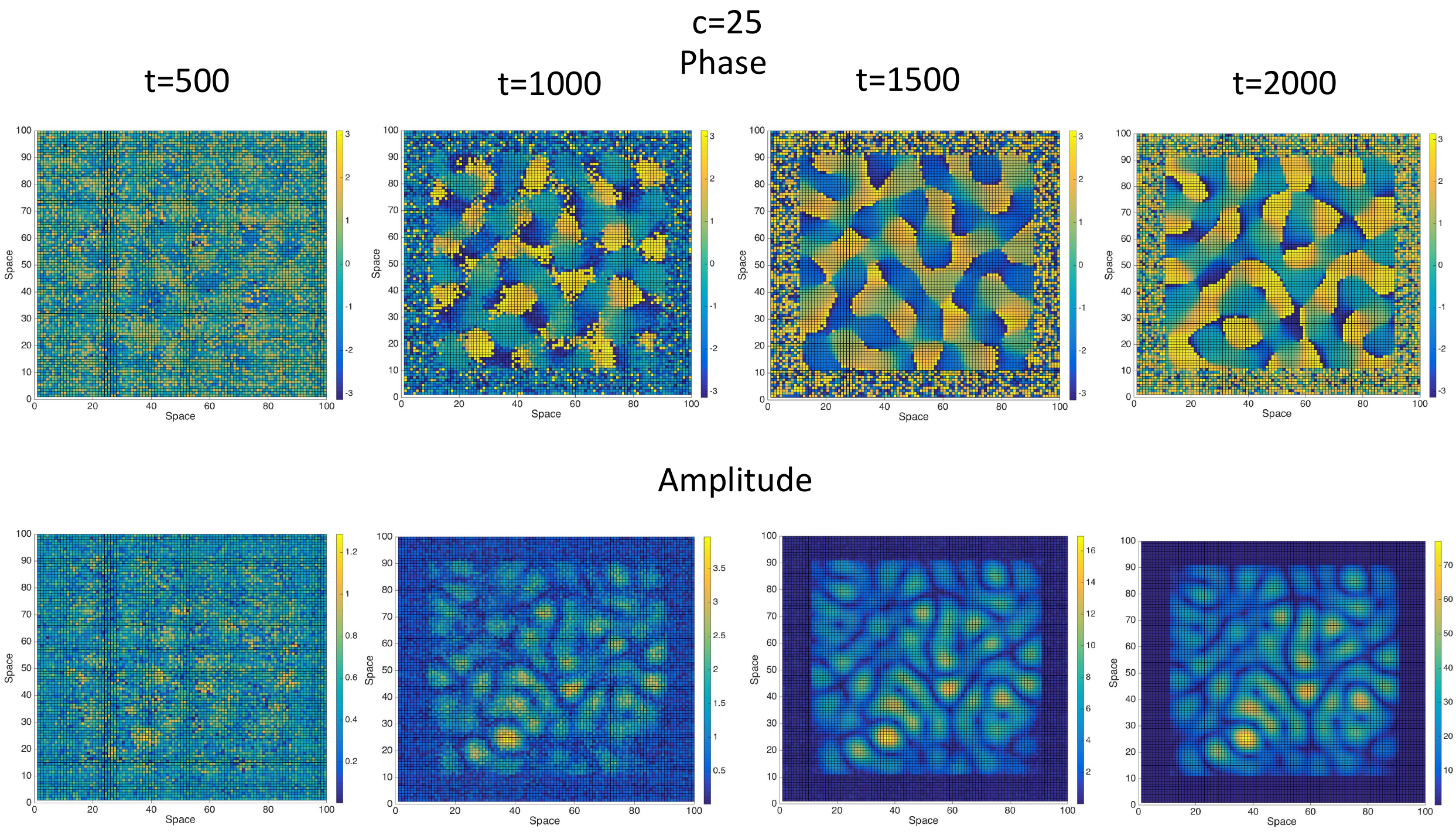} 
\end{center}
\caption{Phases and amplitudes on iterations 500, 1000, 1500, and 2000 of the 10,000, 2D, MH-coupled processes. Parameters were $b_1=1.3, d_2=1.5, c=25$ and those in Table 1.} 
\label{Fig2Dtheta}
\end{figure}

Figure~\ref{Fig2Dtheta} displays typical results on iterations 500, 1000, 1500, and 2000 for a larger value of $c$, viz. $c=25$, and the same values for $b_1, d_2$.  Here, in addition to the spatial ordering of the phases, spatial patterns appear in the amplitudes as well, comprising irregular patches of higher amplitude juxtaposed with others of lower amplitude. Raised patches in the amplitudes roughly correspond to in-phase patches in the phase lattice, and lower amplitudes roughly correspond to edges of the ordered-phase patches. 

Notice that, again, the amplitudes are growing with time because the system is unstable. By iteration 1500 high amplitudes render the noise term of \eqref{Mexphase} very near zero, resulting in deterministic rotation. This result has aspects in common with previous results with somewhat different deterministic models, e.g., \cite{Butler2012} in which Laplacian coupling is used.  Here, however, we focus on the spatio-temporal phases. A major difference is that the processes at fixed locations, $j$, are fluctuations of the potential function $V_j (t)$ (for short durations), rather than firing rates of individual neurons. 

\section{Discussion}\label{discuss}
In \cite{BGW19} we found conditions under which we can expect to see spatial patterns in the values produced by stochastic neural field equations that have only simple damping as a reaction term in the reaction-coupling system. In the present work we derived expressions for the evolution of a stochastic reaction-coupling system in which the reaction parts, with stochasticity but without coupling, do produce quasi-cycle oscillations. We extended the Kuramoto \cite{Kura84} approach to such systems in three ways: (a) our model couples quasi-cycle oscillators instead of limit cycle oscillators, (b) we considered both phase and amplitude, and (c) we coupled the oscillators using a local Mexican Hat (difference-of-Gaussians) coupling instead of all-to-all coupling. It\^o's Lemma produced local couplings of phases and amplitudes in our stochastic system, analogous in form to the couplings described in \cite{Daff11} for a deterministic system.

Global (Kuramoto) and local (Mexican Hat) couplings produce very different results for the evolution of the respective stochastic systems. In particular, sufficiently strong global coupling leads to widespread phase locking at roughly zero phase difference and roughly uniform amplitudes among the local systems \cite{GMWKura}. In the present system \eqref{Mexphase}, \eqref{MexZ}, even weak local coupling leads to local phase ordering in a pattern repeated across space without corresponding amplitude patterns, and stronger local coupling creates corresponding spatial patterns in amplitude. 

Importantly, in the brain, ordering of phases in the manner demonstrated here, without corresponding amplitude patterns, is likely to be useful in enabling synchrony facilitated communciation (e.g., \cite{Fries05, Sejnow10}) with distant brain regions.  For example, information about visual stimuli encoded in V1 firing rate patterns could be sent to V2 via synchronized oscillations without destroying the stimulus information encoded in V1 \cite{Akam14}. In contrast, stronger local coupling, generating phase ordering along with strong amplitude patterns, would cause a brain region to resist change by input from other regions. 

A few of our simulations, for stronger couplings, showed increases in amplitude, although, as we said, we worked in a parameter range, and in a time interval, where this increase was limited. In real neural systems the application of such couplings must be limited to transient responses over limited time, so that the neural field doesn't saturate, and its functionality become compromised. Indeed this is the case in most neural systems \cite{Barth12}; although saturation does occur in some cases (e.g., very intense sensory stimuli) it is generally avoided, and firing rates of neurons are typically rather low and firing is sparsely distributed \cite{Barth12}. Exactly how this is accomplished is still uncertain, but apparently there are cortical mechanisms that promote sparseness \cite{Barth12}. Sparse firing can result in both noisy limit cycles and quasi-cycles \cite{Wallace11}. Moreover, global inhibition tends to destroy synchronization \cite{Terman95} and decrease amplitude, so that, in the presence of certain input, development of a spatial pattern would be inhibited for the duration of the global inhibition. Intermittent application of global inhibition could encourage transient development of varied spatial patterns controlled by parameter values that change with stimulus or other input conditions. Alternatively, changes in local coupling strength could disrupt the development of spatial amplitude patterns. Such conditions could be simulated using our system \eqref{Mexphase}, \eqref{MexZ} along with a global inhibition operator and changing parameter values for the Mexican Hat operator as well as for the values of the synaptic efficacies in \eqref{Kang1}.

\subsection{The inferred long-term phase pattern}
The phase synchronization patterns shown in Figures \ref{Figb1p3d1p5theta} and \ref{Fig2Dtheta} are continuing to evolve even as the overall amplitude continues to grow. These evolving figures suggest that the lines of 2$\pi$ adjustment, in the longer term, will form equidistant parallel lines extending across the 128 processes as in Figure \ref{Figb1p3d1p5c4theta}. One can observe the spatial frequency by counting the number of adjustments at fixed $t$, 7 in Figure \ref{Figb1p3d1p5theta}. 

There appear to be two long-term states of spatial synchrony, one with the lines of $\pi$ phase adjustment going from right to left across the spatial field as in Figure \ref{Figb1p3d1p5theta}, and the other left to right as in Figure \ref{Figb1p3d1p5c4theta} (look at the $t=10,000$ phase plots). The stochastic aspect of the simulated processes shows itself in that the lines of adjustment are not straight.

The slopes of the lines of adjustment are determined by the ratio of the temporal and spatial frequencies, e.g., 70 cycles per second/7 spatial cycles per loop = 10 loops per second in Figures \ref{Figb1p3d1p5theta} and \ref{Figb1p3d1p5c4theta}. A loop here is one transversal of the 128 processes.

It appears in Figure \ref{Figb1p3d1p5c4theta} that in the long term the stationary stochastic process $\theta(j,t)$  will be linear in $j$ and $t$, and that the slope (partial derivative) in each variable will not depend on the other, so that $\theta(j,t)$ is nearly of the form $c_1 j + c_2 t + c_3$. Here $c_1$ is associated with the width of the coupling kernel and $c_2=\omega$, the central frequency of the uncoupled quasi-cycles. If $c_1$ is near zero, the spatial phase is nearly constant in $j$, i.e., the spatial frequency is small.  This will happen when the MH kernel is wide, approaching the Kuramoto case of global coupling.

\subsection{Mexican Hat coupling and noise smoothing in a generic stochastic neural field equation}
In \cite{BGW19} we studied spatial patterns produced by Mexican Hat coupling of stochastic neural field equations with a simple reaction term. We found that without noise the solutions damped to a vanishingly small amplitude on substantial time intervals even in the presence of excitable modes that eventually produced spatial patterns. Added noise amplified, quickly revealed, and sustained these patterns. Moreover, when we used spatially-smoothed noise, the smoothing itself produced spatial patterns that interacted with the patterns produced by Mexican Hat coupling. This result led us to expect that similar noise-smoothing-modulated spatial patterns would be seen in the context of the present stochastic neural field equation that implements a quasi-cycle oscillator as the reaction term. This has yet to be confirmed as we did not study the effects of noise smoothing in the present paper.

\subsection{Comparison to other models}
Our topic in this paper is Mexican Hat coupling of quasi-cycle oscillators. Numerous works in the literature, \cite{Daff11,Faug15,BGW19,Heitman15,Murray89}, among others, describe studies of Mexican Hat coupling but none, to our knowledge, apply the coupling to quasi-cycle oscillators, which, when uncoupled, are driven by noise and die without it. Heitman and Ermentrout \cite{Heitman15} modeled neural activity as a one- and two-dimensional ring of Mexican-hat-coupled phase oscillators, using Kuramoto equations. Their system is like our system \eqref{Mexphase}, but with amplitude $Z=1$ and no noise. Their analyses of stability suggest that their results on spatial patterns as a function of the extent of inhibition in the Mexican Hat operator might be extended to our noise-driven setting.

In the work by Park et al. \cite{Park17}, in a similar setting to that of \cite{Heitman15}, the notion of instantaneous phase response curve is carefully defined and used to derive phase-dynamic equations for coupled deterministic oscillators.  Again, their results might well be extended to coupled quasi-cycle oscillators in a stochastic setting.

Solvable models of deterministic oscillators arranged in a ring with infinite Mexican Hat coupling, and corresponding simulations, appear in \cite{Uezu12, Uezu13}. A paper that extends Kuramoto's result of an attracting invariant manifold for the phases of heterogeneous oscillators to higher dimensional space is \cite{Chandra18}. Extension of these results to quasi-cycle oscillators is certainly feasible.

In another related paper Butler et al. \cite{Butler2012} constructed a 
cortical model from two families of neurons, excitatory and inhibitory, 
each neuron having the states active and quiescent. A pattern of 
connectivity was defined with pairs of neurons forming microcircuits. Neural 
cortex was modeled as a $d$-dimensional lattice where microcircuits are 
connected by writing currents in terms of discrete Laplacian operators 
applied to summary states of the same families of neurons. Equation 
S25 of \cite{Butler2012} appears to play a role similar to our \eqref{Mexphase}, \eqref{MexZ}. In their simulations, under specified parametric conditions, the steady 
state "becomes unstable to spatially inhomogeneous perturbations leading to 
regular pattern formation." Whereas their model was written explicitly in 
terms of operators on states of the two families of neurons, our simpler 
approach directly hypothesizes reaction and coupling terms which might 
result from a variety of explicit derivations. Our parametric exploration 
of patterns of synchrony from a coupled field of quasi-cycle oscillators differs 
markedly from their aims. 

A point of interest is that Butler et al. \cite{Butler2012} argued that the existence of noise-driven patterns seems incompatible with normal visual function. Presumably this is because in their model such patterns fluctuate randomly and would interfere with stimulus input processing. In the system we study, however, such patterns stabilize very quickly, and could interact with or modulate processing of stimulus inputs. A study of our system's response to patterned inputs would clarify this question.

In an earlier related paper, Hutt et al. studied "noise-induced Turing 
transitions in spatial systems" \cite{Hutt08}. Their stochastic integral-differential 
equation (1) plays the role of our \eqref{Mexphase}, \eqref{MexZ}, being even more general. They looked at the power spectrum of the spatial activity at each time point, 
examining separately the spatial Fourier modes defined by a system of equations. 
Stability analysis identified a "Turing phase transition." Specializing to 
Mexican hat coupling, they studied linear stability of their stochastic 
system in terms of modes and corresponding "expansion coefficients," where 
they showed that the first order ones determine linear stability. This result 
lends support to our choice of using the identity in place of their 
sigmoidal $S$ in the coupling term. In \cite{Hutt08} a main point was to compare 
the case of noise that is uncorrelated in space and time (similar to our 
case) with "global fluctuations," noise that is frozen in space, where 
they found the Turing bifurcation threshold is shifted.

Finally, Jung and Mayer-Kress \cite{Jung95} produced a simple space-time firing model in which threshold devices on a 2D lattice are "pulse-coupled", i.e. firing at nearby units at time $t$ produces distance-scaled input to each unit at time $t + \Delta t$, multiplied 
by a coupling constant, $K$. A spontaneous wave starts when $K$ exceeds a 
threshold $K_c$. In the presence of noise, excitatory waves occur for $K < 
K_c$. Classical unimodal stochastic resonance curves of firing as a function 
of noise level are obtained. One would expect a similar result from our 
model \eqref{Kang3} for small $c$ and very large times, required for stationarity. We 
do not pursue this question here.

\begin{appendix}
\section{Details of It\^{o} calculation for Equation \eqref{dY}}
Here we use It\^{o}'s Lemma to express~(\ref{dY}) in polar coordinates in order to obtain \eqref{Mexphase}, \eqref{MexZ} for the Mexican-Hat-coupled system. To do this, for clarity, we first use the simplest form of discrete approximation to the Laplacian operator, the double difference, in place of $\mathbb{M}$. Then in Appendix B we derive \eqref{Mexphase}, \eqref{MexZ} from a more general form of \eqref{dy1}, \eqref{dy2} that represents the Mexican Hat operator.

Writing \eqref{dY} out explicitly for the $j$th stochastic process, where $j$ is the index over the single, discretized, space dimension and the time index $t$ is suppressed, we obtain
\begin{align}\label{dy1}
dy_{1j}=(-\lambda y_{1j} + \omega y_{2j}) dt+(y_{1j-1}-2y_{1j}+y_{1j+1})dt +dW_j^{y_1}
\end{align}
\begin{align}\label{dy2}
dy_{2j}=(-\omega y_{1j} -\lambda y_{2j}) dt+(y_{2j-1}-2y_{2j}+y_{2j+1})dt +dW_j^{y_2}
\end{align}

Let $y_{1j},y_{2j}, j=1,2,...n$ be given by stochastic differential equations, such as \eqref{dy1}, \eqref{dy2}. It\^{o}'s Lemma says that if $f$ is a smooth function on $\mathbb{R}^2$, then
\begin{align}\label{A3}
df\begin{pmatrix}y_{1j}\\y_{2j}\end{pmatrix}=(\nabla f)^\top d\begin{pmatrix}y_{1j}\\y_{2j}\end{pmatrix}+\frac{1}{2}d\begin{pmatrix}y_{1j}\\y_{2j}\end{pmatrix}^\top Hfd\begin{pmatrix}y_{1j}\\y_{2j}\end{pmatrix},
\end{align}
where
\begin{align*}
\nabla f\begin{pmatrix}y_{1j}\\y_{2j}\end{pmatrix}=\bigg(\frac{\partial f}{\partial y_{1j}}, \frac{\partial f}{\partial y_{2j}}\bigg),
\end{align*}
and
\begin{align*}
Hf\begin{pmatrix}y_{1j}\\y_{2j}\end{pmatrix}=\begin{bmatrix}\frac{\partial ^2 f}{\partial y_{1j}^2}& \frac{\partial^2 f}{\partial y_{1j} \partial y_{2j}}\\ \frac{\partial ^2 f}{\partial y_{1j} \partial y_{2j}}& \frac{\partial^2 f}{\partial y_{2j}^2}\end{bmatrix}.
\end{align*}
We wish to compute $df\begin{pmatrix}y_{1j}\\y_{2j}\end{pmatrix}$ and $dg\begin{pmatrix}y_{1j}\\y_{2j}\end{pmatrix}$,
where
\begin{align*}
Z_j=f\begin{pmatrix}y_{1j}\\y_{2j}\end{pmatrix}=(y_{1j}^2+y_{2j}^2)^\frac{1}{2},
\end{align*}
and
\begin{align*}
\theta_j=g\begin{pmatrix}y_{1j}\\y_{2j}\end{pmatrix}=\arctan\bigg(\frac{y_{2j}}{y_{1j}}\bigg),
\end{align*}
and where $y_{1j},y_{2j}$ are defined by \eqref{dy1}, \eqref{dy2}.

We begin by computing $dZ_j$. We find that the gradient is
\begin{align*}
\nabla Z_j\begin{pmatrix}y_{1j}\\y_{2j}\end{pmatrix}=\bigg(\frac{y_{1j}}{Z_j}, \frac{y_{2j}}{Z_j}\bigg),
\end{align*}
and the Hessian is
\begin{align*}
H Z_j\begin{pmatrix}y_{1j}\\y_{2j}\end{pmatrix}=\frac{1}{Z_j}\bigg[I-\frac{1}{Z_j^2}\begin{pmatrix}y_{1j}^2&y_{1j} y_{2j}\\y_{1j} y_{2j}&y_{2j}^2\end{pmatrix}\bigg].
\end{align*}
The first term on the RHS of \eqref{A3} is
\begin{equation}\label{A4}
\begin{split}
& (\nabla Z_j)^\top d\begin{pmatrix}y_{1j}\\y_{2j}\end{pmatrix}=\bigg(\frac{y_{1j}}{Z_j}, \frac{y_{2j}}{Z_j}\bigg) \\
& \begin{pmatrix}(-\lambda y_{1j} + \omega y_{2j}) dt+(y_{1j-1}-2y_{1j} +y_{1j+1})dt+dW_j^{y_1}\\
(-\omega y_{1j} -\lambda y_{2j})dt +(y_{2j-1}-2y_{2j}+y_{2j+1})dt +dW_j^{y_2}\end{pmatrix} \\
& =\frac{1}{Z_j}\bigg(-\lambda(y_{1j}^2+y_{2j}^2)+M_j^Z\bigg)dt+dW_j\\
& \overset{d}=\bigg(-\lambda Z_j+\frac{M_j^Z}{Z_j}\bigg)dt+dW_j,
\end{split}
\end{equation}
where
\begin{align}\label{A5}
M_j^Z=y_{1j} y_{1j-1}-2y_{1j}^2+y_{1j} y_{1j+1}+y_{2j} y_{2j-1}-2y_{2j}^2+y_{2j} y_{2j+1}.
\end{align}
The second term on the RHS of \eqref{A3} is
\begin{align}\label{A6}
\frac{1}{2}d\begin{pmatrix}y_{1j}\\y_{2j}\end{pmatrix}^\top\bigg[\frac{1}{Z_j}\bigg(I-\frac{1}{Z_j^2}\begin{pmatrix}y_{1j}^2&y_{1j} y_{2j}\\y_{1j} y_{2j}&y_{2j}^2\end{pmatrix}\bigg)d\begin{pmatrix}y_{1j}\\y_{2j}\end{pmatrix}\bigg].
\end{align}

First consider the term of \eqref{A6} containing $I$, which gives
\begin{align*}
\frac{1}{2Z_j}\big((dy_{1j})^2+(dy_{2j})^2\big).
\end{align*}
We use \eqref{dy1}, \eqref{dy2}, compute the squares, and obtain several terms. There are two terms containing $(dW_j^{y_1})^2, (dW_j^{y_2})^2$, which we replace by $dt$. All other terms are of lower order. Hence this term yields
\begin{align*}
\frac{dt}{Z_j}.
\end{align*}
Now consider the remaining term of \eqref{A6}. Again we use \eqref{dy1}, \eqref{dy2} to evaluate $dy_{1j}, dy_{2j}$, and again $(dy_{1j})^2=(dy_{2j})^2=dt$. The other terms are all of lower order. The expression reduces to
\begin{align*}
-\frac{1}{2Z_j^3}Z_j^2 dt=-\frac{dt}{2Z_j}.
\end{align*}
The two terms of \eqref{A6} combine to give us
\begin{align*}
\frac{dt}{2Z_j}.
\end{align*}
Combining this with \eqref{A4} we obtain
\begin{align}\label{A7}
dZ_j=\bigg[\bigg(\frac{1}{2Z_j}-\lambda Z_j\bigg)+\frac{M_j^Z}{Z_j}\bigg]dt+dW_j.
\end{align}
Now we compute $M_j^Z$ defined by \eqref{A5} using
\begin{align*}
y_{1j} =Z_j \cos\theta_j,\,\,\, y_{2j} =Z_j \sin\theta_j.
\end{align*}
We obtain
\begin{align*}
%change this below to get rid of v terms
\begin{split}
& M_j^Z=Z_jZ_{j-1}(\cos\theta_j\cos\theta_{j-1})\\
&+Z_jZ_{j+1}(\cos\theta_j\cos\theta_{j+1})-2Z_j^2\\
&=Z_jZ_{j-1}\cos(\theta_j-\theta_{j-1})+Z_jZ_{j+1}\cos(\theta_j-\theta_{j+1})-2Z_j^2.
\end{split}
\end{align*}
With the above for $M_j^Z$, \eqref{A7} becomes
\begin{align}\label{A8}
\begin{split}
& dZ_j=\bigg(\frac{1}{2Z_j}-\lambda Z_j\bigg)dt\\
& +\bigg(Z_{j-1}\cos(\theta_j-\theta_{j-1})+Z_{j+1}\cos(\theta_j-\theta_{j+1})-2Z_j\bigg)dt+dW_j.
\end{split}
\end{align}

The computation of the stochastic differential equation that defines the process $\theta(y_{1j}, y_{2j})=\arctan (y_{2j}/y_{1j})$ goes similarly. First,
\begin{align*}
(\nabla g)^\top=\bigg(\frac{y_{1j}}{Z_j^2},\frac{-y_{2j}}{Z_j^2}\bigg).
\end{align*}
Then
\begin{align*}
\begin{split}
& (\nabla \theta_j)^\top d\begin{pmatrix}y_{1j}\\y_{2j}\end{pmatrix}=\bigg(\frac{y_{1j}}{Z_j^2}, \frac{-y_{2j}}{Z_j^2}\bigg)d\begin{pmatrix}y_{1j}\\y_{2j}\end{pmatrix}\\
& =\frac{1}{Z^2}\bigg(y_{2j}[-\lambda y_{1j}+\omega y_{2j}]-y_{1j}[-\omega y_{1j}-\lambda y_{2j}]\bigg)dt+\frac{M_j^{\theta}dt}{Z_j^2}+\frac{dW_j}{Z_j}\\
& =\frac{1}{Z^2}(\omega y_{2j}^2+\omega y_{1j}^2)dt+\frac{M_j^{\theta}dt}{Z_j^2}+\frac{dW_j}{Z_j}\\
& =\omega dt+\frac{M_j^{\theta}dt}{Z_j^2}+\frac{dW_j}{Z_j},
\end{split}
\end{align*}
where
\begin{align*}
M_j^{\theta}=y_{1j} y_{1j-1}+y_{1j} y_{1j+1}+y_{2j} y_{2j-1}+y_{2j} y_{2j+1}.
\end{align*}
Computing $M_j^{\theta}$ using the definitions of $y_{1j}, y_{2j}$ as for $M_j^Z$ we have
\begin{align*}
\frac{M_j^{\theta}}{Z_j^2}=\frac{Z_{j-1}}{Z_j}\sin(\theta_j-\theta_{j-1})+\frac{Z_{j+1}}{Z_j}\sin(\theta_j-\theta_{j+1}).
\end{align*}
The Hessian
\begin{align*}
H \theta_j=\frac{1}{Z_j^2}\begin{bmatrix}\frac{-2y_{1j} y_{2j}}{Z_j^2}&1-\frac{2y_{1j}}{Z_j^2}\\-1+\frac{2y_{2j}^2}{Z_j^2}&\frac{2y_{1j} y_{2j}}{Z_j^2}\end{bmatrix},
\end{align*}
so the Hessian term in \eqref{A3} with $f=\theta_j$ is
\begin{align*}
\begin{split}
&\frac{1}{2} d\begin{pmatrix}y_{1j}\\y_{2j}\end{pmatrix}^\top\frac{1}{Z_j^2}\begin{bmatrix}-2y_{1j} y_{2j}&1-\frac{2y_{2j}^2}{Z_j^2}\\1+\frac{2y_{1j}^2}{Z_j^2}&\frac{2y_{1j} y_{2j}}{Z_j^2}\end{bmatrix}d\begin{pmatrix}y_{1j}\\y_{2j}\end{pmatrix}\\
& =\frac{1}{2Z_j^2}\bigg[\frac{-2y_{1j} y_{2j}}{Z_j^2}dt+\frac{2y_{1j} y_{2j}}{Z_j^2}dt +l.o.t.\bigg]\approx 0,
\end{split}
\end{align*}
and so the Hessian term does not contribute to $d\theta_j$. Finally,
\begin{align}\label{A9}
\begin{split}
& d\theta_j=\omega dt +\frac{dW_j}{Z_j}\\
& +\bigg(\frac{Z_{j-1}}{Z_j}\sin(\theta_j-\theta_{j-1})+\frac{Z_{j+1}}{Z_j}\sin(\theta_j-\theta_{j+1})\bigg)dt.
\end{split}
\end{align}

\section{Extension to Mexican Hat operator for Equation \eqref{dY}}
We introduce the simplest Mexican Hat operator possible, with the kernel extending over only 2 neighbors on each side of the process of interest and all of the coefficients, including $c$ in \eqref{operator}, noise strength $\mathbb{E}_0$, and coefficients of the Mexican Hat operator, equal to one. It is not difficult to see how this derivation could be extended to different coefficients. Our two stochastic differential equations appear as
\begin{align}\label{B1}
dy_{1j}=(-\lambda y_{1j} + \omega y_{2j})dt+(-y_{1j-2}+y_{1j-1}+y_{1j}+y_{1j+1}-y_{1j+2})dt +dW_j^{y_1}
\end{align}
\begin{align}\label{B2}
dy_{2j}=(-\omega y_{1j} -\lambda y_{2j})dt+(-y_{2j-2}+y_{2j-1}+y_{2j}+y_{2j+1}-y_{2j+2})dt +dW_j^{y_2}
\end{align}
Again, 
\begin{align*}
Z_j=f\begin{pmatrix}y_{1j}\\y_{2j}\end{pmatrix}=(y_{1j}^2+y_{2j}^2)^\frac{1}{2},
\end{align*}
and
\begin{align*}
\theta_j=g\begin{pmatrix}y_{1j}\\y_{2j}\end{pmatrix}=\arctan\bigg(\frac{y_{2j}}{y_{1j}}\bigg).
\end{align*}
Applying It\^o's Lemma as in Appendix A we have
\begin{equation}\label{B3}
\begin{split}
& (\nabla Z_j)^\top d\begin{pmatrix}y_{1j}\\y_{2j}\end{pmatrix}=\bigg(\frac{y_{1j}}{Z_j}, \frac{y_{2j}}{Z_j}\bigg) \\
& \begin{pmatrix}[(-\lambda y_{1j} + \omega y_{2j}) +(-y_{1j-2}+y_{1j-1}+y_{1j}+y_{1j+1}-y_{1j+2})]dt+dW_j^{y_1}\\
[(-\omega y_{1j} -\lambda y_{2j})+(-y_{2j-2}+y_{2j-1}+y_{2j}+y_{2j+1}-y_{2j+2})]dt +dW_j^{y_2}\end{pmatrix} \\& =\frac{1}{Z_j}\bigg(-\lambda(y_{1j}^2+y_{2j}^2)+M_j^Z\bigg)dt+dW_j\\
& \overset{d}=\bigg(-\lambda Z_j+\frac{M_j^Z}{Z_j}\bigg)dt+dW_j,
\end{split}
\end{equation}
as in Appendix A but where here
\begin{align}\label{B4}
\begin{split}
& M_j^Z=y_{1j} (-y_{1j-2}+y_{1j-1}+y_{1j}+y_{1j+1}-y_{1j+2})\\
&+y_{2j} (-y_{2j-2}+y_{2j-1}+y_{2j}+y_{2j+1}-y_{2j+2}).
\end{split}
\end{align}
The remainder of It\^o's Lemma gives the same results as in Appendix A for any $dy_{1j}$ and $dy_{2j}$, so we have for the Mexican Hat coupling:
\begin{align}\label{B5}
dZ_j=\bigg(\frac{1}{2Z_j}-\lambda Z_j\bigg)dt+\frac{M_j^Z}{Z_j}dt+dW_j.
\end{align}
Notice that \eqref{B5} is only different from \eqref{A7} in the form of $M_j^Z$. This means that when we insert $y_{1j}=Z_j\cos\theta_j$ and $y_{2j}=Z_j\sin\theta_j$ into $M_j^Z$ to obtain the final form, no matter what coupling we use, we just have to compute $M_j^Z$ and then compute the result of the insertion, because only $M_j^Z$ has $y_{1j}$ and $y_{2j}$ terms in it.

Now, inserting the polar coordinate expressions for $y_{1j}$ and $y_{2j}$ into \eqref{B4}, we have
\begin{align*}
\begin{split}
& M_j^Z=-Z_j\cos\theta_j Z_{j-2}\cos\theta_{j-2} +Z_j\cos\theta_j Z_{j-1}\cos\theta_{j-1}+(Z_j\cos\theta_j )^2\\
& +Z_j\cos\theta_j Z_{j+1}\cos\theta_{j+1}-Z_j\cos\theta_j Z_{j+2}\cos\theta_{j+2}\\
&-Z_j\sin\theta_j Z_{j-2}\sin\theta_{j-2} +Z_j\sin\theta_j Z_{j-1}\sin\theta_{j-1}+(Z_j\sin\theta_j )^2\\
& +Z_j\sin\theta_j Z_{j+1}\sin\theta_{j+1}-Z_j\sin\theta_j Z_{j+2}\sin\theta_{j+2}.
\end{split}
\end{align*}
Collecting terms yields
\begin{align*}
\begin{split}
& M_j^Z=Z_j\big[-Z_{j-2}\cos(\theta_j-\theta_{j-2})+Z_{j-1}\cos(\theta_j-\theta_{j-1})\\
& +Z_{j+1}\cos(\theta_j-\theta_{j+1})-Z_{j+2}\cos(\theta_j-\theta_{j+2})+Z_j\big]\\
& =Z_j\bigg[\sum_{i=-N}^{N}\mathbb{M}_{ji}Z_{j+i}\cos(\theta_j-\theta_{j+i})\bigg],
\end{split}
\end{align*}
where $\mathbb{M}_{ji}$ represents the coefficients of the Mexican Hat operator. The final expression for the approximation with the just-derived coupling becomes
\begin{align}\label{B6}
dZ_j=\bigg(\frac{1}{2Z_j}-\lambda Z_j\bigg)dt+\bigg(\sum_{i=-N}^{N}\mathbb{M}_{j,j+i}Z_{j+i}\cos(\theta_j-\theta_{j+i})\bigg)dt+dW_j.
\end{align}

The derivation of the more general form for $d\theta_j$ proceeds in the same fashion, with the result that
\begin{align}\label{B7}
d\theta_j=\omega dt +\bigg(\sum_{i=-N}^{N}\mathbb{M}_{j,j+i}\frac{Z_{j+i}}{Z_j}\sin(\theta_j-\theta_{j+i})\bigg)dt+\frac{db_j}{Z_j}.
\end{align}

\end{appendix}

\section*{Competing interests}
The authors declare that they have no competing interests.
\section*{Author's contributions}
Both authors contributed to the conceptualization and writing of the paper. The numerical simulations were accomplished by LMW.

\medskip
% The data information below will be filled by AIMS editorial staff
Received xxxx 20xx; revised xxxx 20xx.
\medskip

\end{document}